\documentstyle[12pt,epsfig,rotating]{article}
\def\npb#1#2#3{{\it Nucl.\ Phys.\ }{\bf B #1} (#2) #3}
\def\pl#1#2#3{           {Phys. Lett. }{\bf #1} (19#2) #3}

\def\prd#1#2#3{{\it Phys.\ Rev.\ }{\bf D #1} (#2) #3}

\def\gsim{\raise0.3ex\hbox{$\;>$\kern-0.75em\raise-1.1ex\hbox{$\sim\;$}}}

\newcommand {\fig} [1] {Fig.~\ref{#1}}

\begin{document} 
\begin{titlepage} 
\begin{flushright}
hep-ph/0307364 \\ 
IFIC/03-40\\
ZU-TH 11/03 \\ 
\end{flushright} 
\vspace*{3mm} 
\begin{center}  
  \textbf{\large Neutrino properties and the decay of the lightest 
supersymmetric particle}\\[10mm]

{M. Hirsch${}^1$ and  W. Porod${}^2$}
\vspace{0.3cm}\\ 

{\it $^1$ Astroparticle and High Energy Physics Group, 
IFIC - Instituto de F\'\i sica Corpuscular,
Edificio Institutos de Investigaci\'on,
Apartado de Correos 22085,
E-46071 Valencia, Espa\~na \\}
{\it $^2$ Institut f\"ur Theoretische Physik, Universit\"at Z\"urich, \\
CH-8057 Z\"urich, Switzerland}

\end{center}

\begin{abstract} 
Supersymmetry with broken R-parity can explain the neutrino mass squared 
differences and mixing angles observed in neutrino oscillation experiments. 
In the minimal model, where R-parity is broken only by bilinear terms, 
certain decay properties of the lightest supersymmetric particle (LSP) are 
correlated with neutrino mixing angles. Here we consider charginos, squarks, 
gluinos and sneutrinos being the LSP and calculate their decay properties 
in bilinear R-parity breaking supersymmetry. Together with the decays of 
charged scalars and neutralinos calculated previously this completes the 
proof that bilinear R-parity breaking as the source of neutrino masses will 
be testable at future colliders. Moreover, we argue that in case of GMSB,
the decays of the NLSP can be used to test the model.
\end{abstract} 
 
\end{titlepage}

\newpage

\setcounter{page}{1} 

\section{Introduction}

In supersymmetric models with R-parity violation (RpV) 
\cite{BRpV,Hall:1983id,TRpV} the lightest supersymmetric particle (LSP) 
is unstable and decays. Thus astrophysical constraints on its nature 
\cite{Dimopoulos:1989hk} no longer apply and a priori {\em any} SUSY 
particle could be the LSP. On the other hand, most studies of LSP 
phenomenology at accelerators in RpV models up to now have concentrated 
on a) the lightest neutralino \cite{NtrlDecay,NtrlOthers} or b) a charged 
scalar (most probably the right scalar tau) \cite{ChScDecay,ChOthers} 
being the LSP. The purpose of the present work is to study the phenomenology 
of the remaining LSP candidates (charginos, gluinos, scalar quarks and 
scalar neutrinos) at future colliders for the case where R-parity is 
broken by bilinear terms only.

Arguably at present the main motivation to study RpV SUSY models is the 
astonishing {\em experimental} progress in neutrino physics in the past 
few years. Super-K observations of atmospheric neutrinos \cite{Fukuda:1998mi}, 
solar neutrino measurements by the SNO collaboration \cite{Ahmad:2002jz} and 
the reactor anti-neutrino experiment KamLAND \cite{Eguchi:2002dm} have 
finally established non-zero neutrino masses and mixings beyond any 
reasonable doubt. Bilinear RpV (BRpV) SUSY models necessarily produce 
Majorana neutrino masses and indeed in \cite{NuMass} it was shown that 
BRpV SUSY can explain current neutrino data, once 1-loop corrections are 
carefully taken into account.

Can one test BRpV SUSY being the origin of neutrino masses and mixings? 
Conventional wisdoms says, if SUSY is to solve the gauge hierarchy 
problem, superpartners should be found at the next generation of 
colliders. In general, the only predictable difference between R-parity 
conserving and R-parity violating SUSY then would be a decaying LSP. 
Given the relatively small number of free parameters in BRpV SUSY, however, 
one can go further and from measured data on neutrino properties {\em predict} 
several decay properties of the LSP. This was shown for the case of the 
neutralino being LSP in \cite{NtrlDecay} and for charged scalar LSPs 
in \cite{ChScDecay}. In both cases, interesting relations 
between certain decay patterns and low-energy neutrino data have been 
found. The most important are: (i) The ratio 
$BR(\chi^0_1 \to \mu q \bar{q}')/BR(\chi^0_1 \to \tau q \bar{q}') \simeq
 \tan^2 \theta_{Atm}$ and (ii) 
$BR(\tilde \tau_1 \to e \nu)/BR(\tilde \tau_1 \to \mu \nu) \simeq
 \tan^2 \theta_{sol}$. 

The minimal supersymmetric extension of the standard model (MSSM) 
\cite{Haber:1984rc} contains more than a hundred free parameters, 
most of which are soft SUSY breaking masses and phases. 
Supplementing the MSSM with universal mSugra boundary conditions
\cite{Haber:1984rc} reduces this number to just four free parameters 
plus an undetermined sign (in addition to the standard model parameters). 
These are given at the grand unification (GUT) scale as: $m_0$, the 
common scalar mass, $m_{1/2}$, the gaugino mass, and $A_0$, the common 
trilinear parameter. In addition one usually chooses $\tan\beta =v_u/v_d$ 
and the sign of the Higgs mixing term $|\mu|$ as free parameters. 
Parameters at the electro-weak scale can then be calculated from RGE 
running, reducing considerably the available parameter space. 

In such a constrained version of the MSSM (CMSSM) one finds essentially 
only two LSP candidates, namely, the lightest neutralino and the right 
sleptons, in particular the right scalar tau if $\tan \beta$ is large. 
Arguably it is this theoretical prejudice why also in RpV models other 
possibilities so far have attracted little attention.

Models which depart from strict mSugra in one way or another, however, 
can be found in the literature. Just to mention a few representative 
examples, there are string inspired models where supersymmetry breaking
is triggered not only by the dilaton fields but also by moduli fields
\cite{Brignole:1993dj}. In $SO(10)$ or $E(6)$ models, where all neutral
gauge bosons, except those forming $Z$ and $\gamma$, have masses 
of the order of $m_{GUT}$ one expects additional D-term contributions
to the sfermion mass parameters at $m_{GUT}$ \cite{Kolda:1995iw}. 
This is equivalent to assuming non-universal values of $m_0$ for 
left-sleptons, right-sleptons, left-squarks and right squarks, giving 
rise e.g.~to sneutrino LSPs. In gauge mediated supersymmetry breaking (GMSB) 
models \cite{gmsb2} exists the possibility, that 
 the gluino is the LSP \cite{Raby:1997bp}. In AMSB models 
\cite{Giudice:1998xp} one can find parameter regions where the chargino 
is the LSP (nearly mass degenerate with the lightest neutralino). 

To be as general as possible, however, in this work we will not resort 
to any specific model of SUSY breaking. Instead we will simply point 
out in which way one has to depart from mSugra to obtain the corresponding 
LSP and then proceed to calculate its decay properties. To summarize 
our main result it can be said that independent of which SUSY particle is 
the LSP there is at least one ratio of decay branching ratios which 
is fixed by either the solar or the atmospheric neutrino angle, i.e. 
independent of the LSP nature, BRpV SUSY as the origin of neutrino masses 
is testable at future colliders.

This paper is organized as follows. In the next section we will recapitulate 
the main features of the bilinear R-parity violating model. Then we will turn 
to the numerical calculations. Decays of squarks, gluinos, charginos and 
scalar neutrinos will be discussed in detail. Before concluding with a short 
summary, we will argue that also in case of GMSB \cite{Giudice:1998xp},  
BRpV SUSY remains testable due to the decay patterns of the NLSP.

\section{The Model}

Bilinear R-parity breaking supersymmetry has been discussed extensively 
in the literature \cite{BRpV,NuMass}. We will therefore summarize only 
the main features of the model here, with emphasis on neutrino physics. 
The Lagrangian of the model is obtained by adding bilinear
terms breaking lepton number to the MSSM  superpotential:
\begin{equation}
W_{\rm{BRpV}} = W_{\rm{MSSM}}  - \varepsilon_{ab}
\epsilon_i \widehat L_i^a\widehat H_u^b \; ,
\end{equation}
and consistently the corresponding terms to the soft SUSY breaking potential:
\begin{equation}
V_{\rm{soft}} = V_{\rm{soft,MSSM}} -\varepsilon_{ab}
          B_i\epsilon_i\widetilde L_i^aH_u^b \, .
\end{equation}
The latter induce vacuum expectation values $v_i$ for the sneutrinos which
are in turn responsible for mixing between standard model particles with
supersymmetric particles: Higgs bosons with sleptons, charged leptons with
charginos and, most importantly for the following considerations, neutrinos
with neutralinos. 

The mixing of neutrinos with neutralinos gives rise to one massive neutrino
at tree level. The other two neutrinos obtain masses due to loop
effects \cite{NuMass}. Assuming that the heaviest neutrino obtains its 
mass at tree level, the main features relevant for our current purpose are 
the following.  The mass of the heaviest neutrino is given
by:
\begin{eqnarray}
m_{\nu_3} &=& \frac{(g M_1 + g' M_2) |\vec \Lambda|^2}
                 {4 {\rm Det}(\tilde \chi^0)} \\
 \Lambda_i &=& \epsilon_i v_d + \mu v_i \, .
\end{eqnarray}
The atmospheric neutrino mixing angle is given by
\begin{eqnarray}
 \tan \theta_{\rm Atm} = \left|\frac{\Lambda_2}{\Lambda_3}\right|
\end{eqnarray}
and the so-called CHOOZ angle by
\begin{eqnarray}
U_{e3}^2 \simeq \frac{\Lambda_1^2}{\Lambda_2^2 + \Lambda_3^2} \, .
\end{eqnarray}
The scale of the loop-induced solar mass is given by  
\begin{eqnarray}
m_{\nu_2} \propto \frac{|\vec \epsilon|^2}{16 \pi^2 \mu^2} m_b 
\end{eqnarray}
and the solar mixing angle by
\begin{eqnarray}
 \tan \theta_{\rm sol} &\simeq&
          \left|\frac{\tilde \epsilon_1}{ \tilde \epsilon_2}\right| \\
  \tilde \epsilon_i &=& V^{\nu,\rm tree}_{ij} \epsilon_j 
\end{eqnarray}
where  $V^{\nu,\rm tree}$ is the tree level neutrino mixing 
matrix \cite{NuMass}. In the region where the condition 
$(\epsilon_2 \Lambda_2)/(\epsilon_3 \Lambda_3) <0$ is fulfilled one finds
that $\tilde \epsilon_1/ \tilde \epsilon_2 \simeq \epsilon_1/ \epsilon_2$.
For a more thorough discussion see ref.~\cite{NuMass}. 

In this model the neutrino spectrum is hierarchical and hence the
neutrino mass scales coincide with the (square roots of) the mass squared 
differences measured in oscillation experiments.  
This implies that the R-parity violating parameters are significantly
smaller than the R-parity conserving parameters: 
$|\epsilon_i| \ll |\mu$ and $|v_i| \ll v_d$. This feature allows for
the possibility that all R-parity violating couplings can be expanded
in terms of the ratios
\begin{eqnarray}
 \frac{\epsilon_i}{\mu}, \, \,
 \frac{\Lambda_i}{\sqrt{{\rm Det} (\tilde \chi^0)}} \hskip2mm \rm{or}\hskip2mm
 \frac{\Lambda_i}{{\rm Det}(\tilde \chi^+)} \, .
\end{eqnarray}
Several examples of this kind can be found in 
\cite{NtrlDecay,ChScDecay,NuMass,Restrepo:2001me}.
We have used this possibility for a systematic expansion of the R-parity 
violating couplings to obtain a semi-analytical understanding of the 
results presented in the following section.
For completeness we want to note that in our model 
also gauginos and gauge bosons have R-parity violating couplings implying
that one can clearly distinguish BRpV from a model where only trilinear
R-parity violating couplings are present.

\section{Numerical results}

In this section we present various collider observables and their 
correlations with neutrino observables for the following LSP candidates: 
charginos, sneutrinos, squarks and gluino. The numerical results
are obtained in the following way except if stated otherwise:
(i) We create a random sample over the SUSY parameter space, 
using five free parameters: $m_0$, $m_{1/2}$, $A_0$, $tan\beta$ and 
$\mu$. Motivated by mSugra, we calculate the gaugino masses 
approximately from $m_{1/2}$, for the sfermion mass parameters 
we use $m_0$ directly at the electroweak scale. We have checked 
explicitly for several LSP sets that the latter simplification 
has no impact on our results. We then violate one of the following 
mSugra conditions at a time. 
Each condition is necessary but not sufficient to obtain the 
corresponding LSP, i.e. after calculating the SUSY spectrum we 
post-select points which have the desired candidate LSP. 
Chargino LSPs are obtained by the condition 
$m_2 \simeq (5/3) \tan^2\theta_W m_1$, sneutrino LSPs 
by $m_{L_i}^2 \ll m_{0}^2$, squark LSPs by $m_Q^2,m_D^2,m_U^2 \ll m_0^2$ 
and gluino LSPs by $m_3 \ll m_2,m_1$.
(ii) The R-parity violating parameters are added such, that
$\Delta^2_{\rm Atm}$ and  $\Delta^2_{\rm sol}$ are consistent with 
the experimental data.

We want to stress that the correlations between low-energy and high-energy 
observables shown in the following are predictions after a generous sampling 
over the SUSY parameter space. Much tighter correlations could be 
obtained, once information on at least a part of the SUSY
spectrum is put in, see e.g.~ref.~\cite{NtrlDecay} for the case of
neutralinos. In particular information on $\tilde \chi^0_j$, $\tilde
\chi^+_k$, $\tilde \tau_i$, $\tilde b_i$ and $H^+$ would be 
important in this respect.

\subsection{Charginos}

For chargino LSPs possible final states are
\begin{eqnarray}
 \tilde \chi^+_1 &\to&  \sum_{q=d,s;q'=u,c} \bar{q} q' \nu_i;
                        \bar{b} t \nu_i  \\
 \tilde \chi^+_1 &\to& l^+_i \sum_{q=u,d,s} q \bar{q} ;
                      l^+_i \bar{c} c; l^+_i \bar{b} b; l^+_i \bar{t} t;     \\
 \tilde \chi^+_1 &\to& l^+_i l^+_j l^-_k  \\
 \tilde \chi^+_1 &\to& l^+_i \nu_r \nu_s 
\end{eqnarray}
 where $l_i = e,\mu,\tau$ and we sum over the 
three neutrino flavours as well as the lighter quark states 
$u$, $d$ and $s$ which cannot be separated experimentally. Note, that
we calculate here the 3-body decays even if an intermediate real
2-body final state is possible by including the finite width of the
intermediate states. These intermediate states contain in general a gauge
boson, whose R-parity violating couplings to  charginos are
typically an order of magnitude smaller compared to the R-parity
violating couplings of the virtual sfermions.

Numerically one finds that the final state $\bar{q} q' \sum_j \nu_j$ has 
usually the largest branching ratio (up to 65 \%). Typical branching 
ratios for other final states not containing a top quark are in 
the range of several per mille to a few per--cent. Final states with 
top quarks are found to be always very small or kinematically closed, 
because $m_{\tilde \chi^{+}_1}$ does not exceed 300 GeV in our numerical 
data sets. Moreover, the intermediate states are always off-shell in
this case contrary to final states involving light quarks.

\begin{figure}
\setlength{\unitlength}{1mm}
\begin{center}
\begin{picture}(80,50)
\put(0,-30){\mbox{\epsfig{figure=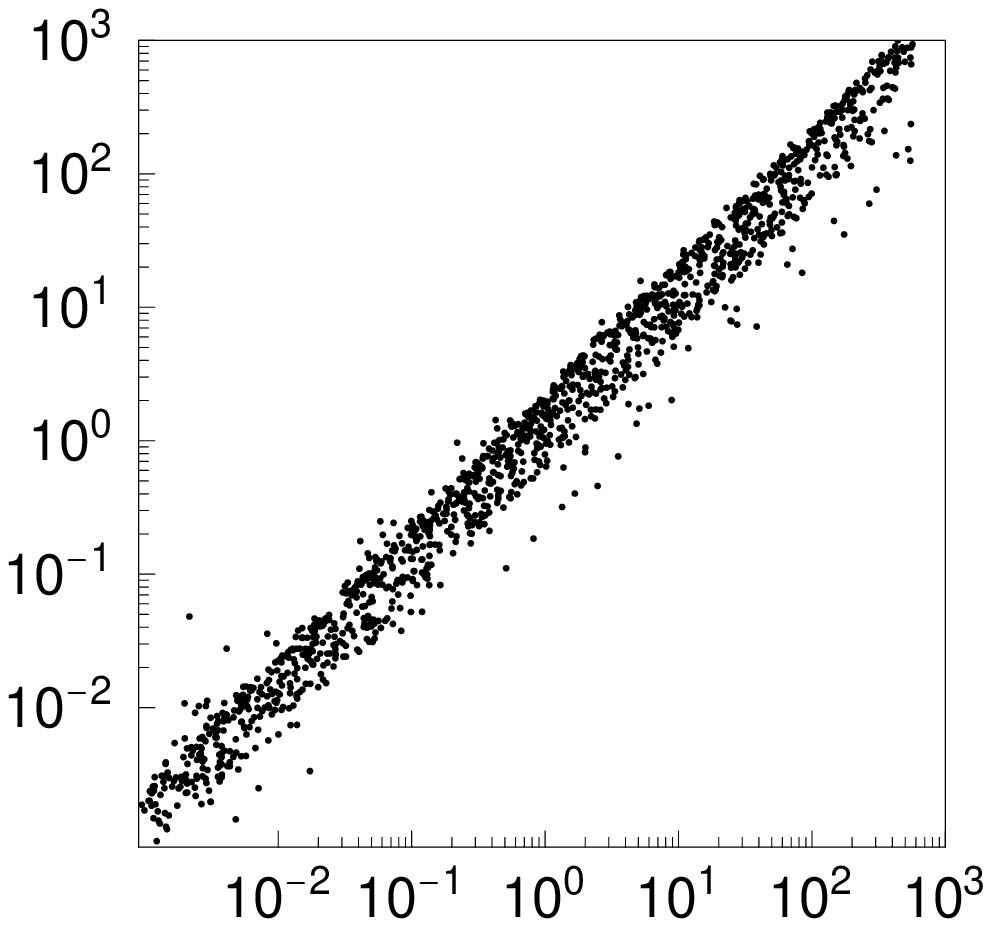,height=7.7cm,width=7.cm}}}
 \begin{rotate}{90}
    \hskip-2mm {{\small $\Gamma/m_{\tilde \chi^+}^5$ [meV/(100 GeV)$^5$]}}
\end{rotate}
\put(71,-33){\makebox(0,0)[br]{{$m_{\nu_3}$ [eV]}}}
\end{picture}
\end{center}
\vskip28mm
\caption[]{Chargino decay width divided by $m_{\tilde \chi^+}^5$ as a function 
of the heaviest neutrino mass}
\label{fig:ChDecayLength}
\end{figure}

As a first example how neutrino physics allows to predict observables 
for collider physics we plot in \fig{fig:ChDecayLength}
$\Gamma/m_{\tilde \chi^+}^5$ [meV/(100 GeV)$^5$] as a function of the 
heaviest neutrino mass $m_{\nu_3}$ [eV]. We scale out a fifth power 
of the chargino mass to account approximately for the phase space 
of the decay. \footnote{Although this was not discussed in 
\cite{NtrlDecay} a similar correlation holds for the case of 
the lightest neutralino being the LSP.}

\fig{fig:ChDecayLength} shows an obvious correlation between chargino 
decay width and neutrino mass. However, there is a sizable spread in the 
prediction. Thus such a measurement could probably never compete with 
neutrino oscillation experiments in terms of accuracy. Nevertheless, from 
the current data on atmospheric neutrinos one can roughly predict,
\begin{equation}
\label{eq:PredGamCh}
\frac{\Gamma}{m_{\tilde \chi^+}^5} = (0.02 - 1.2)\hskip5mm 
              [\frac{meV}{(100 GeV)^5}]
\end{equation}
Eq.(\ref{eq:PredGamCh}) can be considered as a consistency check for the 
completeness and uniqueness of the bilinear model as the main source 
of the (atmospheric) neutrino mass. Significantly smaller or larger 
widths would be a clear signal that BRpV cannot explain the neutrino 
data. Note that the width of the band gets reduced sizeably once some 
information on the SUSY spectrum is available. 

\begin{figure}
\hskip2mm
\epsfysize=60mm
\epsfxsize=60mm
\epsfbox{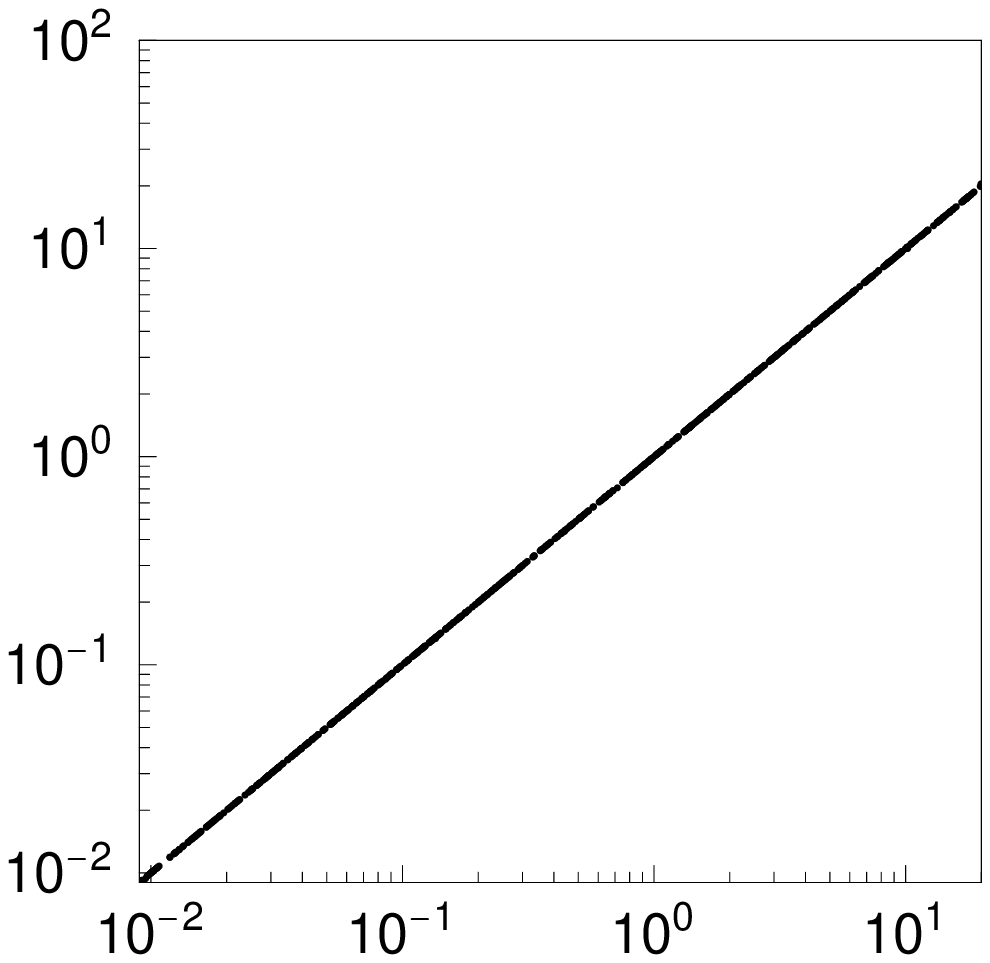}

\vskip-60mm
\hskip72mm
\epsfysize=60mm
\epsfxsize=60mm
\epsfbox{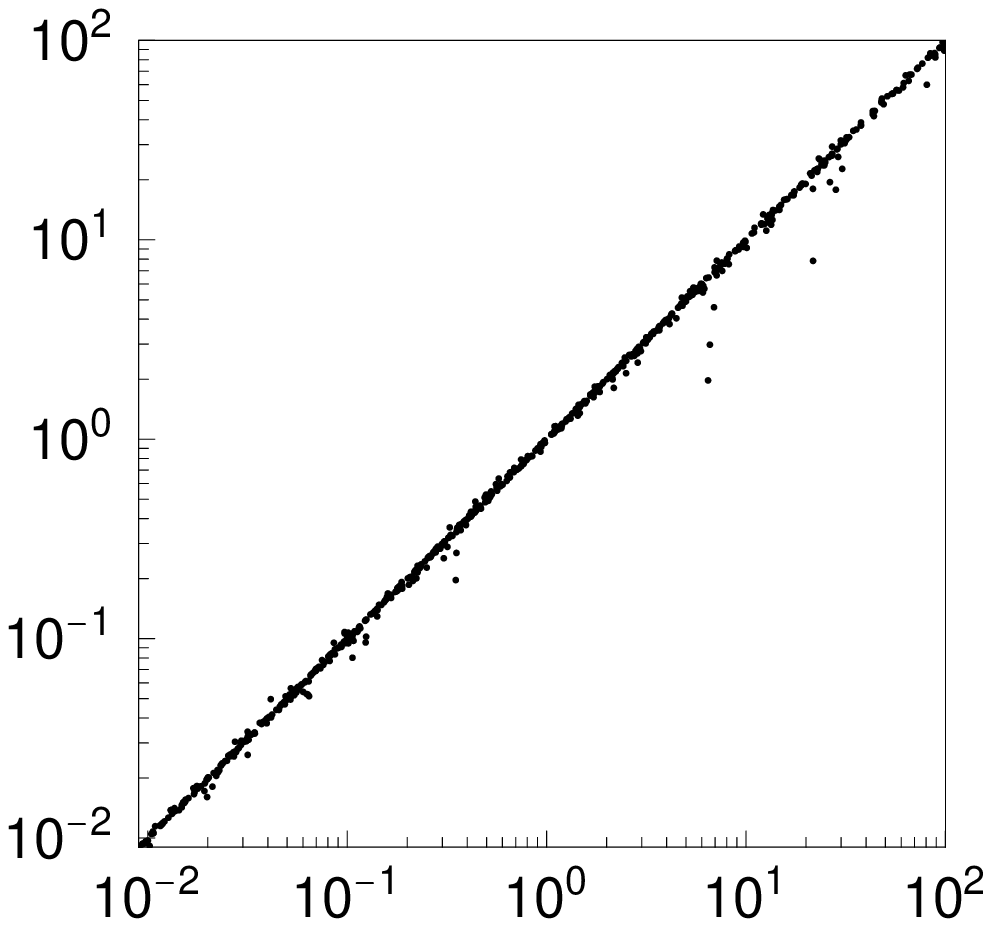}

\vskip-10mm
\begin{rotate}{90}
$BR({\tilde \chi}^{+} \rightarrow e {\bar c}c)/
BR({\tilde \chi}^{+} \rightarrow \mu {\bar c}c)$ 
\end{rotate}

\vskip-5mm
\hskip70mm
\begin{rotate}{90}
$BR({\tilde \chi}^{+} \rightarrow \mu {\bar c}c)/
BR({\tilde \chi}^{+} \rightarrow \tau {\bar c}c)$ 
\end{rotate}

\vskip5mm
\hskip45mm
$(\Lambda_1/\Lambda_2)^2$

\vskip-5mm
\hskip110mm
$(\Lambda_2/\Lambda_3)^2$

\caption[]{Ratio of branching ratios for chargino decay. To the left, 
$BR({\tilde \chi}^{+} \rightarrow e {\bar c}c)/
BR({\tilde \chi}^{+} \rightarrow \mu {\bar c}c)$ 
as a function of $(\Lambda_1/\Lambda_2)^2$. To the right, 
$BR({\tilde \chi}^{+} \rightarrow \mu {\bar c}c)/
BR({\tilde \chi}^{+} \rightarrow \tau {\bar c}c)$ 
as a function of $(\Lambda_2/\Lambda_3)^2$.}
\label{fig:ChCC}
\end{figure}

As discussed previously \cite{NtrlDecay}, ratios of different branching ratios 
of the LSP decays can trace information on ratios of RpV parameters. 
In case of the lightest chargino being the LSP we have found that 
various ratios are sensitive to ratios of $\Lambda_i/\Lambda_j$.
Two examples are shown in \fig{fig:ChCC}. In this figure 
we show $BR({\tilde \chi}^{+} \rightarrow e {\bar c}c)/
BR({\tilde \chi}^{+} \rightarrow \mu {\bar c}c)$ (to the left)  
and $BR({\tilde \chi}^{+} \rightarrow \mu {\bar c}c)/
BR({\tilde \chi}^{+} \rightarrow \tau {\bar c}c)$ (to the right) as function 
of $\Lambda_1/\Lambda_2$ and $\Lambda_2/\Lambda_3$, respectively. 
Obviously measurements of these branching ratios would determine 
the corresponding ratio of $\Lambda$'s to high accuracy. Somewhat worse
results are obtained if one has to sum over the quarks of the
first two generations. 

\begin{figure}
\setlength{\unitlength}{1mm}
\begin{center}
\begin{picture}(80,50)
\put(0,-30){\mbox{
   \epsfig{figure=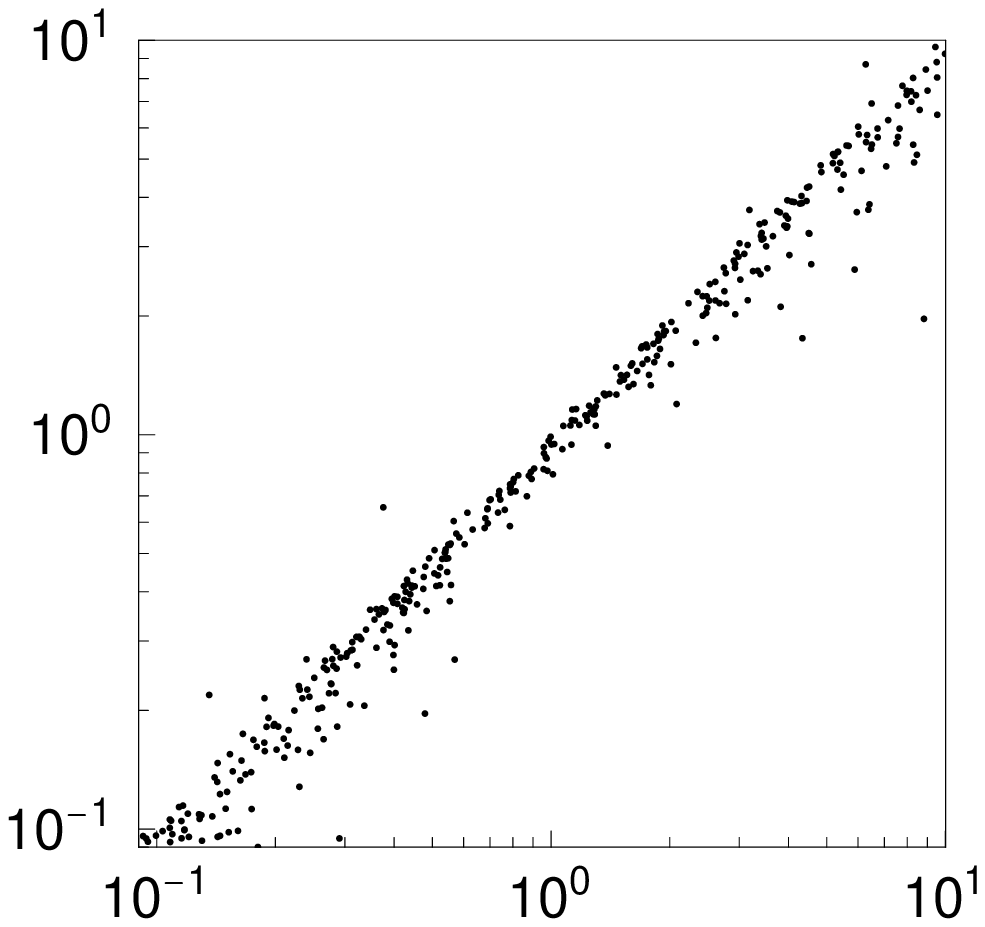,height=7.7cm,width=7.cm}}}
\begin{rotate}{90}
    \hskip-2mm {{\small $BR(\tilde \chi^+ \rightarrow \mu qq)/BR(\tilde \chi^+ 
\rightarrow \tau q q)$}}
\end{rotate}
\put(71,-33){\makebox(0,0)[br]{{$\tan^2\theta_{Atm}$}}}
\end{picture}
\end{center}
\vskip28mm
\caption[]{Ratio of chargino decay branching ratios 
$BR(\tilde \chi^+ \rightarrow \mu qq)/BR(\tilde \chi^+ \rightarrow \tau q q)$
versus $\tan^2\theta_{Atm}$. For the currently preferred value of 
$\tan^2\theta_{Atm}=1$ one expect approximately equal branching ratios 
for these final states.}
\label{fig:ChvAtm}
\end{figure}

Since $\Lambda_2/\Lambda_3$ determines the atmospheric angle in BRpV, 
one expects the ratios discussed above to be correlated with 
$\tan^2\theta_{Atm}$. That this is indeed the case is shown in 
\fig{fig:ChvAtm}, where we plot 
$BR({\tilde \chi}^{+} \rightarrow \mu {\bar q}q)/
BR({\tilde \chi}^{+} \rightarrow \tau {\bar q}q)$ as a function of 
$\tan^2 \theta_{\rm Atm}$ summing over all quarks of the first
two generations. For the currently preferred value of 
$\tan^2\theta_{Atm}=1$ one expect approximately equal branching ratios 
for these final states, nearly independent of any other parameter. 

Due to the fact that the chargino has many possible final states, 
one can devise various cross checks of the bilinear model. We have 
found that the following ratios are sensitive to $\Lambda_1/\Lambda_2$ 
only to a good approximation: $BR({\tilde \chi}^{+} \rightarrow e {\bar q}q)/
BR({\tilde \chi}^{+} \rightarrow \mu {\bar q}q)$, 
$BR({\tilde \chi}^{+} \rightarrow 3 e)/
 BR({\tilde \chi}^{+} \rightarrow \mu 2 e)$ and 
$BR({\tilde \chi}^{+} \rightarrow 3 e)/
 BR({\tilde \chi}^{+} \rightarrow 3 \mu)$. 

Similarly there are a number of ratios which trace very well the 
ratio $\Lambda_2/\Lambda_3$. Since the latter quantity is related 
to the atmospheric angle, using the currently available experimental 
data on $\tan^2(\theta_{Atm})$, one can {\em predict} various ratios 
of branching ratios. Ranges allowed by current data are listed in Table 
\ref{tab:TestAtm}. 

\begin{table}
\caption[]{Ratio of branching ratios for chargino LSP decays as required 
by the consistency of
the model. These ratios all trace the ratio $\Lambda_2/\Lambda_3$. The 
experimentally allowed range for the atmospheric neutrino angle, 
$0.3 \le \sin^2(\theta_{Atm})\le 0.7$ (at 3 $\sigma$ c.l.), has been 
used to obtain the quoted ranges. Ratios have been sorted with respect 
to increasing uncertainties.}
\begin{center}
\begin{tabular}{|l|c|c|} \hline
  Ratio & lower bound & upper bound  \\ \hline 
Br($ \mu \bar{c} c$) / Br($ \tau \bar{c} c$) &
                                                     0.45 & 1.4 \\
Br($ \mu \bar{q} q$) / Br($ \tau \bar{q} q$) &
                                                     0.45 & 1.4 \\
Br($ \mu 2 e$) / Br($ \tau 2 e$) &
                                                     0.45 & 1.4 \\
Br($ \mu \bar{b} b$) / Br($ \tau \bar{b} b$) &
                                                     0.46 & 2.1 \\
Br($ 3 \mu$) / Br($ \tau 2 \mu$) &
                                                     0.28 & 1.8 \\
Br($ 3 \mu $) / Br($ 3\tau $) &
                                                     0.096 & 1.4 \\
\hline
\end{tabular}
\end{center}
\label{tab:TestAtm}
\end{table}

\subsection{Sneutrinos}

In the case that sneutrinos are the LSPs they
will decay according to
\begin{eqnarray}
 \tilde \nu_i &\to& q \, \bar{q}  \\
 \tilde \nu_i &\to& l^+_j \, l^-_k \\
 \tilde \nu_i &\to& \nu_j \, \nu_k 
\end{eqnarray}

\begin{figure}
\setlength{\unitlength}{1mm}
\begin{center}
\begin{picture}(160,180)
%
%
\put(5,145){\mbox{\epsfig{
              figure=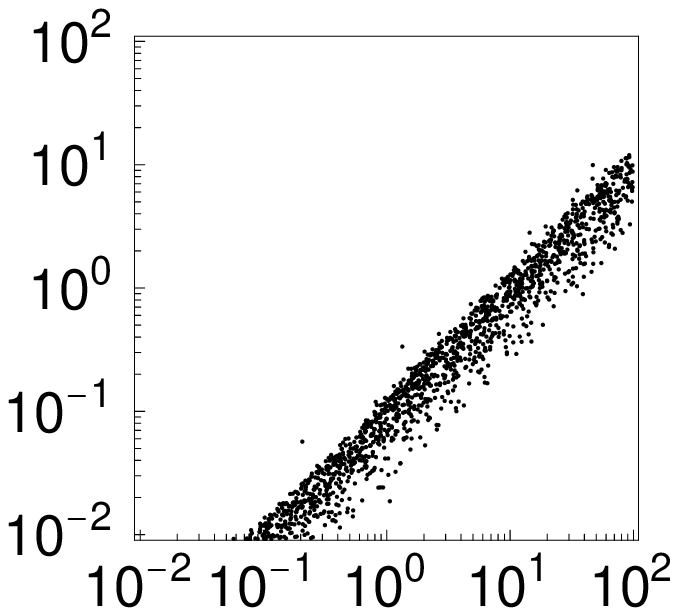,height=6.5cm,width=7cm}}}
\put(3,165){\makebox(0,0)[br]{
    \begin{rotate}{90}
           {{\small $\sigma * BR$($e \mu$) /$\sigma * BR$($\mu \mu$) }}
    \end{rotate}}}
\put(59,145){\mbox{{\small $(\epsilon_1/\epsilon_2)^2$ }}}
\put(22,199){\mbox{{\small (a) }}}
\put(88,144){\mbox{\epsfig{
            figure=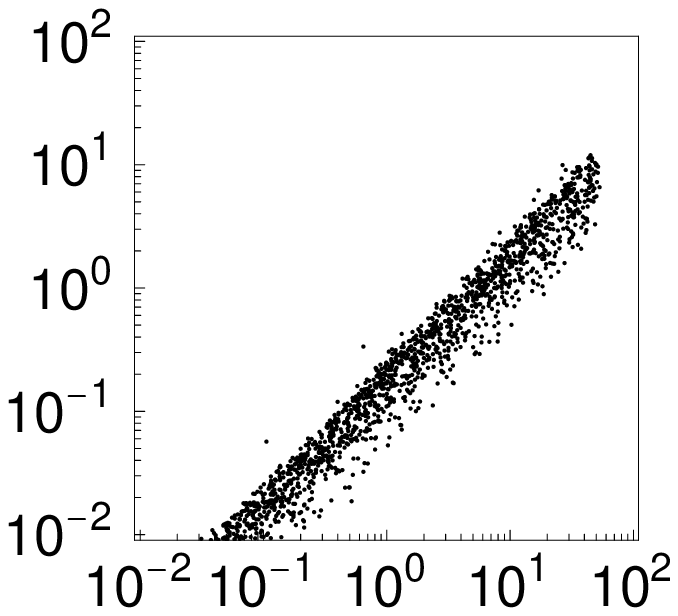,height=6.5cm,width=7cm}}}
\put(86,165){\makebox(0,0)[br]{
    \begin{rotate}{90}
         {{\small $\sigma * BR$($e \mu$) /$\sigma * BR$($\mu \mu$) }}
    \end{rotate}}}
\put(152,145){\makebox(0,0)[br]{{\small $\tan^2\theta_{sol}$}}}
\put(105,198){\mbox{{\small (b) }}}
%
%
\put(5,70){\mbox{\epsfig{
              figure=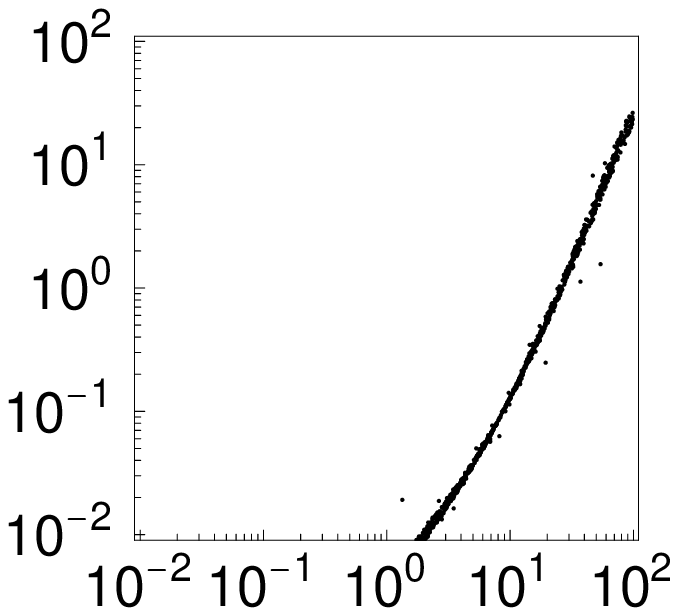,height=6.5cm,width=7cm}}}
\put(3,85){\makebox(0,0)[br]{
    \begin{rotate}{90}
        {{\small  $\sigma * BR$($e \mu$) /$\sigma * BR$($\mu \tau$) }}
    \end{rotate}}}
\put(59,69){\mbox{{\small $(\epsilon_1/\epsilon_2)^2$ }}}
\put(22,124){\mbox{{\small (c) }}}
\put(88,70){\mbox{\epsfig{
            figure=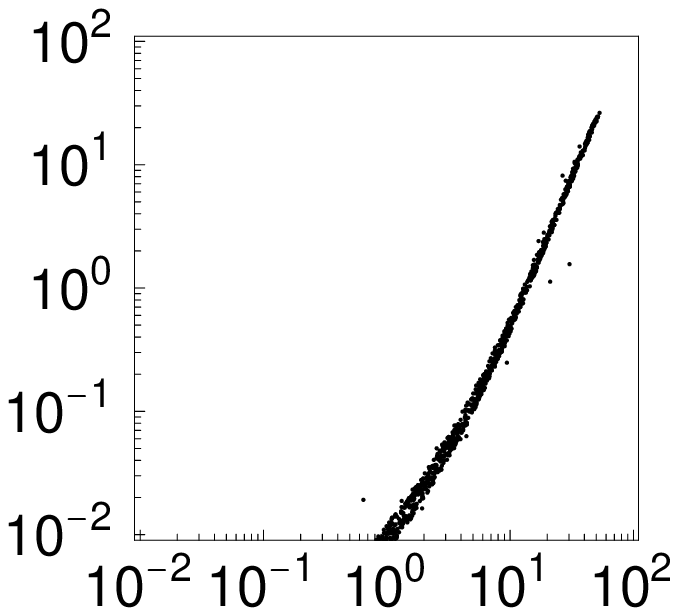,height=6.5cm,width=7cm}}}
\put(86,85){\makebox(0,0)[br]{
    \begin{rotate}{90}
            {{\small   $\sigma * BR$($e \mu$) /$\sigma * BR$($\mu \tau$) }}
    \end{rotate}}}
\put(152,69){\makebox(0,0)[br]{{\small $\tan^2\theta_{sol}$}}}
\put(105,124){\mbox{{\small (d) }}}
%
%
\put(5,-5){\mbox{\epsfig{
          figure=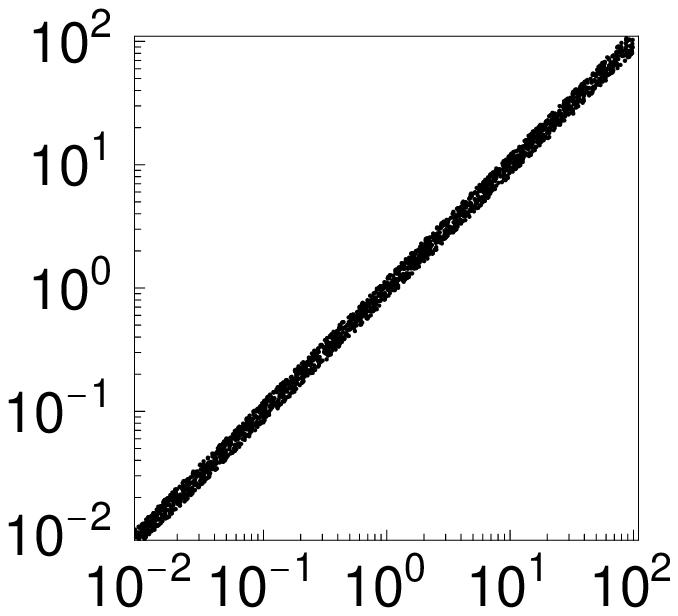,height=6.5cm,width=7cm}}}
\put(3,0){\makebox(0,0)[br]{
    \begin{rotate}{90}
        {{\small BR($\tilde \nu_\mu \to e^\pm \mu^\mp$)
                /BR($\tilde \nu_\mu \to \mu^\pm \tau^\mp$)}}
    \end{rotate}}}
\put(59,-6){\mbox{{\small $(\epsilon_1/\epsilon_2)^2$}}}
\put(22,49){\mbox{{\small (e) }}}
\put(88,-5){\mbox{\epsfig{
    figure=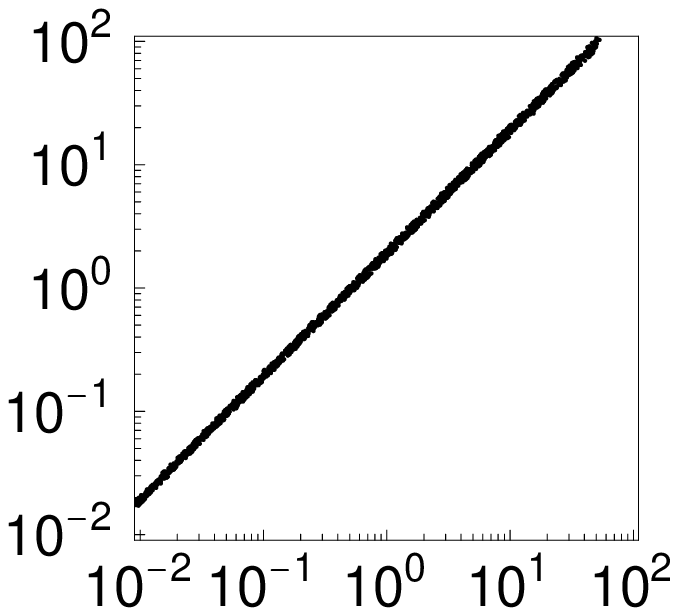,height=6.5cm,width=7cm}}}
\put(86,0){\makebox(0,0)[br]{
    \begin{rotate}{90}
        {{\small BR($\tilde \nu_\mu \to e^\pm \mu^\mp$)
                /BR($\tilde \nu_\mu \to \mu^\pm \tau^\mp$) }}
    \end{rotate}}}
\put(152,-6){\makebox(0,0)[br]{{\small $\tan^2\theta_{sol}$}}}
\put(105,49){\mbox{{\small (f) }}}
\end{picture}
\end{center}
\caption[]{Various sneutrino observables as a function of 
           $(\epsilon_1/\epsilon_2)^2$ (left column) and $\tan^2\theta_{sol}$ 
           (right column). 
           $\sigma * BR(l_i l_j)$ is defined as 
           $\sum_{r,s=1}^3 \sigma(e^+ e^- \to \tilde \nu_s \tilde \nu_r
     \to b \bar{b} l^\pm_i l^\mp_j )$. For discussion see text. }
\label{fig:sneutrino}
\end{figure}

Scanning over the parameter space one finds the following general features:
(i) The main decay mode is for  all sneutrinos $\tilde \nu_i \to b \bar{b}$. 
(ii) The branching ratios for decays into charged leptons are O($10^{-2}$). 
(iii) The invisible decay mode into two neutrinos is O($10^{-3}$) and below. 
Moreover, it turns out that the decay lengths of all sneutrinos have the
same order of magnitude because (a) the largest couplings are of the form
$\epsilon_i Y_b$ and (b) the largeness of the solar mixing angle 
requires $\epsilon_1 \simeq \epsilon_2$ and one expects that also 
$\epsilon_3$ is of similar size. This implies that one has to
sum over all sneutrinos if one considers the direct production at an
$e^+ e^-$ collider. 
Therefore we consider the following quantity:
\begin{eqnarray}
 \sigma*BR(l^\pm_i l^\mp_j) \equiv
  \sum_{r,s=1}^3 \sigma(e^+ e^- \to \tilde \nu_s \tilde \nu_r
     \to b \bar{b} l^\pm_i l^\mp_j ) 
\end{eqnarray}
In \fig{fig:sneutrino}a -- d we show the ratio of these observables as
a function of $(\epsilon_1/\epsilon_2)^2$ and $\tan^2 \theta_{\rm
sol}$.  The correlations of the collider observables with neutrino physics 
are obvious. The range for the
observables under study is $\sigma * BR$($e \mu$) =
O($10^{-2})$--O(1)~fb, $\sigma * BR$($\mu \mu$) = O(0.1)~fb, 
$\sigma * BR$($\mu \tau$) = O(0.1)--O(10)~fb at a 800 GeV $e^+ e^-$ linear
collider and unpolarized beams. Larger cross sections are obtained for 
polarized beams with left--handed electrons and right--handed
positrons.  We want to stress again, that the spread in the
correlations is mainly due to the unknown SUSY spectrum. It gets
considerably reduced once information on the SUSY spectrum is plugged
in.

In principal one could tag the flavour of the sneutrino in
cascade decays, e.g.~in the decay $\tilde \chi^+_1 \to l^+ \tilde \nu$
and then study the subsequent decay of the sneutrino into leptons.
Due to the fact that all involved decays are two-body decays,
one can distinguish the lepton stemming from the chargino
from the ones stemming from the sneutrino by measuring the lepton
energy except in the case where 
$m_{\tilde \chi^+_1} \simeq 3 m_{\tilde \nu} /2$. In \fig{fig:sneutrino}e 
and f we show correlations between the branching ratio of the muon sneutrino
and  $(\epsilon_1/\epsilon_2)^2$ and $\tan^2 \theta_{\rm sol}$
assuming that such a flavour tag can indeed be performed.
As can be seen, current results for the solar mixing angle predict
that BR($\tilde \nu_\mu \to e^\pm \mu^\mp$)
                /BR($\tilde \nu_\mu \to \mu^\pm \tau^\mp)$ is 
in the range $0.4-2$ 
independent of the remaining SUSY parameters.

In scenarios, where sneutrinos are the LSPs, the left charged sleptons 
are not much heavier. The difference between the masses of the charged left
sleptons and sneutrinos is roughly given by: 
$m^2_{\tilde l,L} - m^2_{\tilde \nu} \simeq  - \cos 2 \beta m^2_W > 0$.
Depending on the mass difference the left sleptons decay in these scenarios
either via three body decays, which conserve R-parity, into 
\cite{Ambrosanio:1997bq,Djouadi:2000bx}
\begin{eqnarray}
 \tilde l_L &\to& \tilde \nu \, \bar{q} q' \\
 \tilde l_L &\to& \tilde \nu \, \nu l       
\end{eqnarray}
or via R-parity violating couplings into
\begin{eqnarray}
 \tilde l_L &\to&  \bar{q} q' \\
 \tilde l_L &\to&  \nu l \, .       
\end{eqnarray}
The latter decay modes give in principal rise to additional observables 
correlated with neutrino physics. However, we have found that for mass 
differences larger
than $\simeq 5$ GeV the three body decays clearly dominate. Therefore this 
additional information is only aviable if either $\tan\beta$ is small and/or
if all particles have masses above $\gsim 400$~GeV.

\subsection{Squarks}

Here we discuss the decays of the squarks of the first two
generations as well as the decays of the lighter sbottom.
The decays of the lighter stops have been discussed in detail in 
ref.~\cite{Restrepo:2001me}
and are  similar to the results for the sbottoms discussed below.

In the case that squarks are the lightest SUSY particle, they will decay
according to
\begin{eqnarray}
 \tilde q &\to& q \, \nu \\
 \tilde q &\to& q' \, l
\end{eqnarray}
In case that the mass difference between the squarks of the 
first two generation is larger than approx.
5 GeV, the heavier ones will dominantly decay according to
\begin{eqnarray}
\tilde q &\to& \tilde q' \, q    \, \bar{q}'
\end{eqnarray}
mediated mainly via gluino exchange. Here we have used
the formulas given in \cite{Djouadi:2000bx}. This decay mode dominates
the decays of $\tilde d_L$, because $m_{\tilde d_L} > m_{\tilde u_L} +
5$~GeV for $m_{\tilde Q} < 500$~GeV and $\tan\beta\ge 3$ due to the
D-terms in the mass matrix.

\begin{figure}
\setlength{\unitlength}{1mm}
\begin{center}
\begin{picture}(160,180)
%
%
\put(5,120){\mbox{\epsfig{figure=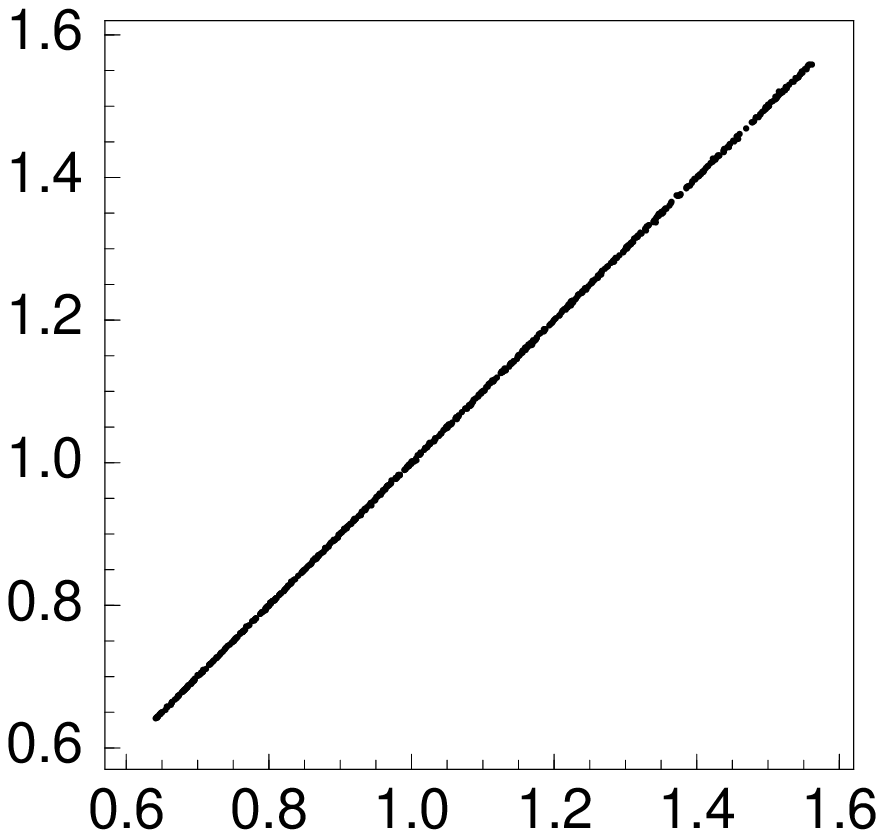,height=6.5cm,width=7cm}}}
\put(3,130){\makebox(0,0)[br]{
    \begin{rotate}{90}
         {{\small BR($\tilde u_L \to l^+_i \bar{d}$)
                 /BR($\tilde u_L \to l^+_j \bar{d}$) }}
    \end{rotate}}}
\put(59,119){\mbox{{\small $(\epsilon_i/\epsilon_j)^2$ }}}
\put(22,173){\mbox{{\small (a) }}}
\put(88,120){\mbox{\epsfig{figure=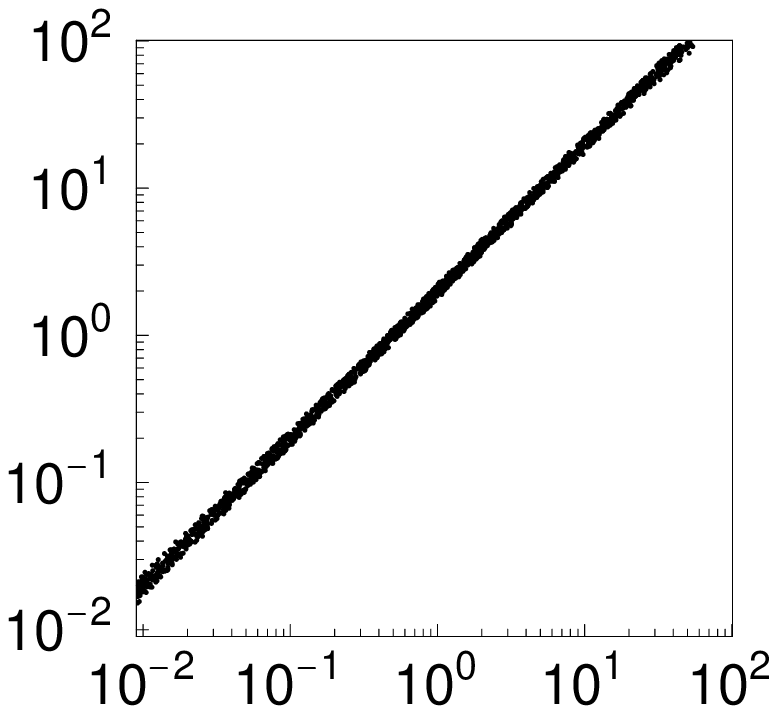,height=6.5cm,width=7cm}}}
\put(86,130){\makebox(0,0)[br]{
    \begin{rotate}{90}
       {{\small BR($\tilde u_L \to e^+ \bar{d}$)
               /BR($\tilde u_L \to \mu^+ \bar{d}$) }}
    \end{rotate}}}
\put(138,119){\mbox{{\small $(\tan \theta_{\rm sol})^2$ }}}
\put(105,172){\mbox{{\small (b) }}}
%
%
\put(5,45){\mbox{\epsfig{figure=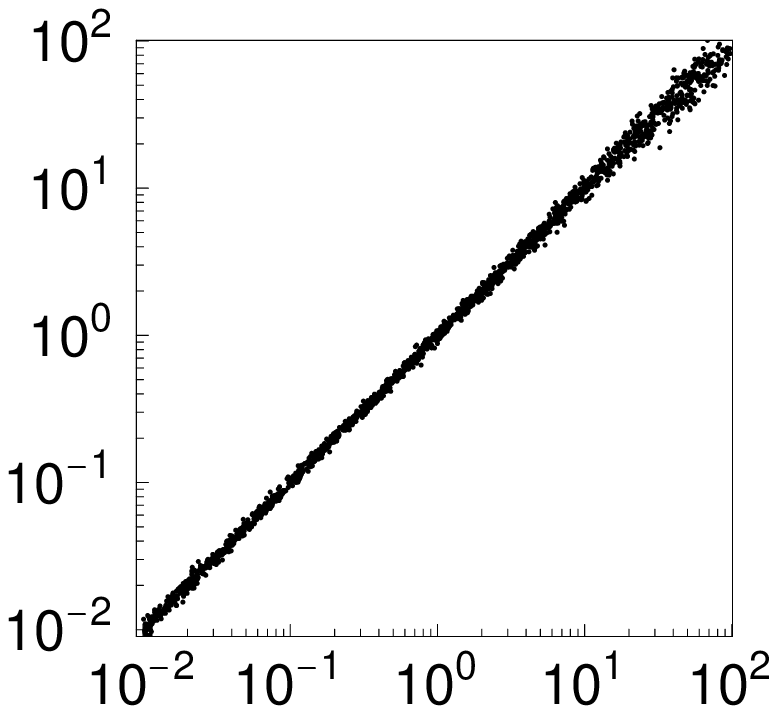,height=6.5cm,width=7cm}}}
\put(3,55){\makebox(0,0)[br]{
    \begin{rotate}{90}
         {{\small BR($\tilde q \to l_i^\pm \bar{q}'$)
                /BR($\tilde q \to l_j^\pm \bar{q}'$) }}
    \end{rotate}}}
\put(59,44){\mbox{{ $(\epsilon_i/\epsilon_j)^2$}}}
\put(22,99){\mbox{{\small (c) }}}
\put(88,45){\mbox{\epsfig{figure=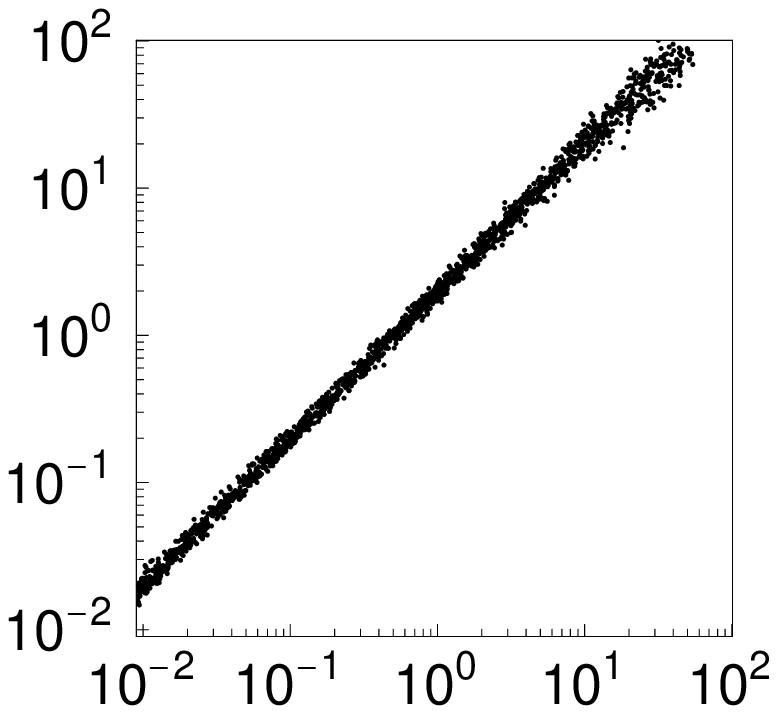,height=6.5cm,width=7cm}}}
\put(86,55){\makebox(0,0)[br]{
    \begin{rotate}{90}
              {{ \small BR($\tilde q \to e^\pm q'$)
                     /BR($\tilde q \to \mu^\pm q'$) }}
    \end{rotate}}}
\put(138,44){\mbox{{\small $(\tan \theta_{\rm sol})^2$ }}}
\put(105,99){\mbox{{\small (d) }}}
%
%
\put(5,-30){\mbox{\epsfig{figure=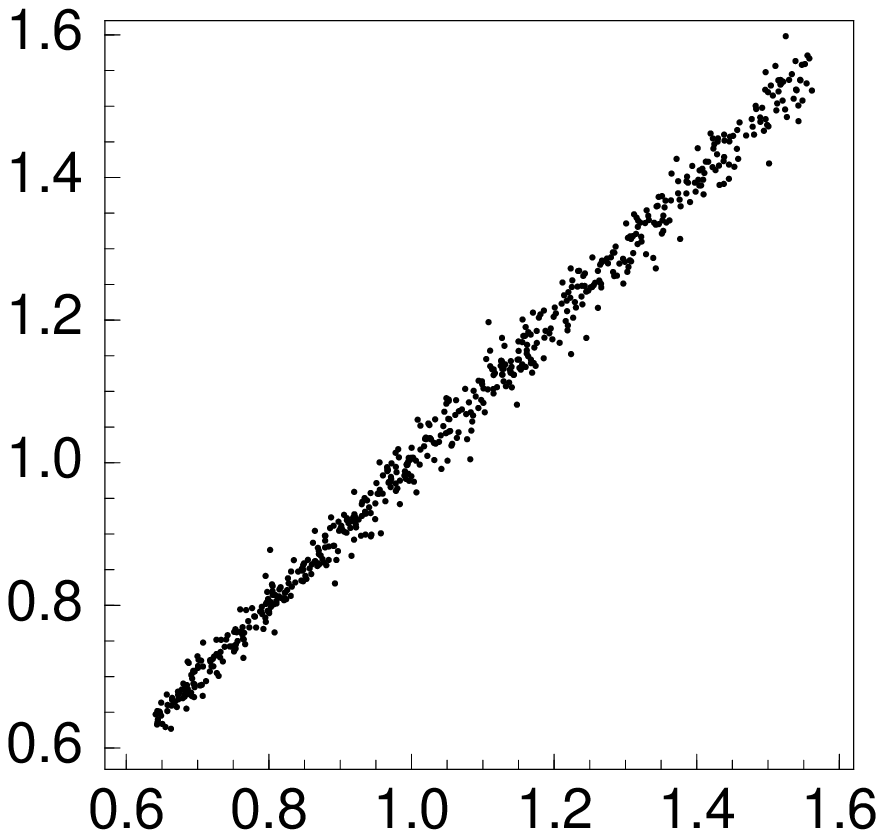,height=6.5cm,width=7cm}}}
\put(3,-20){\makebox(0,0)[br]{
    \begin{rotate}{90}
      {{\small BR($\tilde b_1 \to l^+_i \bar{t}$)
              /BR($\tilde b_1 \to l^+_j \bar{t}$)}}
    \end{rotate}}}
\put(59,-31){\mbox{{ $(\epsilon_i/\epsilon_j)^2$}}}
\put(22,23){\mbox{{\small (e) }}}
\put(88,-30){\mbox{\epsfig{figure=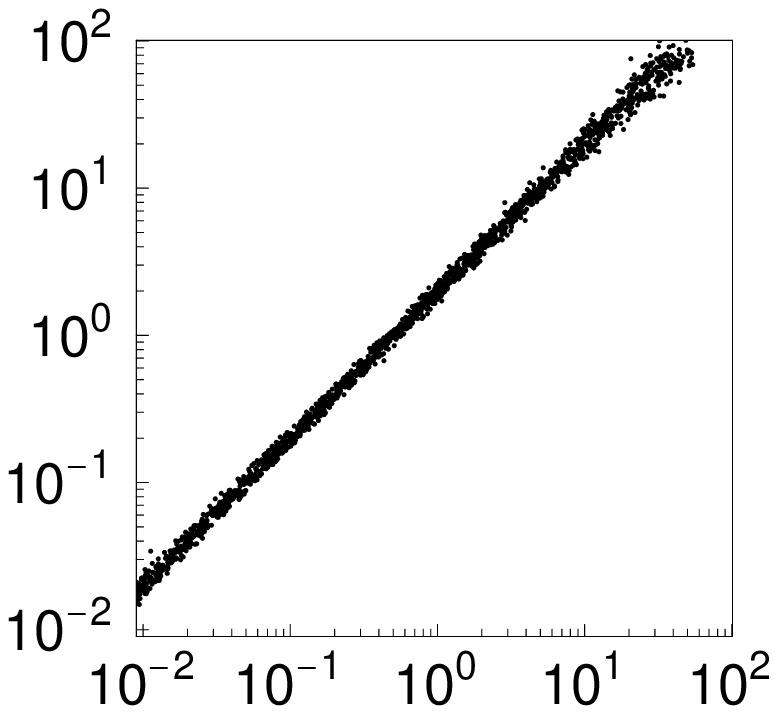,height=6.5cm,width=7cm}}}
\put(86,-20){\makebox(0,0)[br]{
    \begin{rotate}{90}
      {{\small BR($\tilde b_1 \to e^+ \bar{t}$)
              /BR($\tilde b_1 \to \mu^+ \bar{t}$)}}
    \end{rotate}}}
\put(138,-31){\mbox{{ $(\tan \theta_{\rm sol})^2$ }}}
\put(105,23){\mbox{{\small (f) }}}
%
%
%
\end{picture}
\end{center}
\vskip28mm
\caption[]{Ratios of branching ratios for squark decays as 
               function of $(\epsilon_i/\epsilon_j)^2$ (left column) 
               and $\tan^2\theta_{sol}$ (right column). }
\label{fig:squarks}
\end{figure}

The important decay modes related to neutrino physics are those induced by 
the effective
$\hat L_i \hat Q \hat D^c_R$ coupling which is proportional to
$\epsilon_i h_d$. In the numerical results below we assume that
the first two generations of squarks are mass degenerate. This 
assumption is motivated by the experimental constraints from meson physics, 
in particular for the $K^0$-$\bar{K}^0$ mixing \cite{Gabbiani:1996hi}.
 However,
we do not assume left and right squarks have the same mass.

Typical examples are shown in \fig{fig:squarks}, where we consider
three different scenarios: 
(i) In \fig{fig:squarks}a and b we show the
case where  $\tilde u_L$ and   $\tilde c_L$ are the LSPs.  
One sees that the corresponding relations between $\tan^2 \theta_{sol}$ 
(or $(\epsilon_1/\epsilon_2)^2$) and ratios of branching ratios into charged
leptons are extremely pronounced. 
Note that the decays into leptons clearly dominate, 
with a total branching ratio of 0.6 -- 0.9 summed over all charged 
leptons.
(ii) In \fig{fig:squarks}c and d we show how the results are changed if
one sums over the left and right squarks of the first two
generations. As discussed above, $\tilde d_L$ and $\tilde s_L$ are not
included in this sum because their R-parity violating decay modes
are negligible. The correlations are somewhat worse compared to the previous
case because $\tilde u_R$ ($\tilde c_R$) decay into leptons only
via their mixing with the corresponding left partner and the corresponding
branching ratios are of order $10^{-3}$--$10^{-2}$. The
branching ratios of $\tilde d_R$ and $\tilde s_R$ into leptons is 
approximately in the range 0.05 - 0.5.
(iii) In \fig{fig:squarks}e and f we show the correlations if 
$\tilde b_1$ is the LSP. One sees that the ratio of branching ratios
into top quarks is nicely related to the ratios of the $\epsilon_i$
squared. In the case that this decay is suppressed the decays into a lepton
and a $c$--quark give similar, although somewhat worse, results. In this
case the branching ratio is of order $10^{-2}$.

Finally, we want to note that these correlations are hardly affected by 
QCD corrections because they nearly drop out 
by taking the ratio of branching ratios. We have checked this
explicitly by adopting the formulas given in \cite{Kraml:1996kz} to
the bilinear model.

\subsection{Gluinos}

\begin{figure}
\setlength{\unitlength}{1mm}
\begin{center}
\begin{picture}(80,60)
%
%
\put(5,-3){\mbox{
   \epsfig{figure=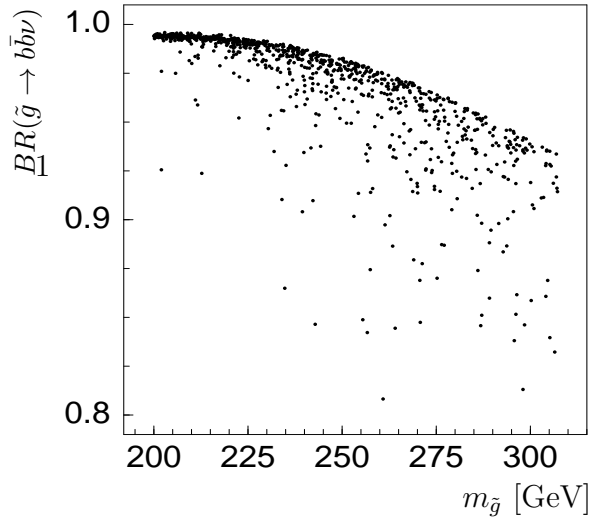,height=6.5cm,width=7.cm}}}
\put(3,38){\makebox(0,0)[br]{
    \begin{rotate}{90}
       {{\small $BR( \tilde g \rightarrow b \bar{b} \nu )$}}
    \end{rotate}}1}
\put(77,-6){\makebox(0,0)[br]{{$m_{\tilde g}$~[GeV]}}}
\end{picture}
\end{center}
\caption[]{ $BR( \tilde g \rightarrow b \bar{b} \nu )$ as a function
of $m_{\tilde g}$}
\label{fig:gluinoa}
\end{figure}

In the case that the gluino is the LSP, it decays according to
\begin{eqnarray}
\tilde g &\to& \nu_i q \bar{q} \\
\tilde g  &\to& l^\pm q q' \\
\tilde g  &\to& \nu_i g
\end{eqnarray}
These decays proceed via virtual squarks. For this reason one
expects also correlations between ratios of branching ratios into 
$l_i q \bar{q}'$ and ratios of
 $\epsilon_i$ and, thus, the solar mixing angle. 
We have adopted the formulas given in \cite{Bartl:bu} for the calculation.
The general features of a gluino LSP are:
(i) The decay into the final state  $\nu_i b \bar{b}$ dominates, 
where we sum over all neutrinos. This can be seen  in  \fig{fig:gluinoa}.
(ii) The sum of the branching ratios of the decays into 
$l^\pm b t$ ($l=e,\mu,\tau$) final states is of order 
$10^{-2}$--$10^{-1}$ if summed over all charged leptons.  
Note that at the LHC
O($10^5$) gluino pairs can be produced per year if $m_{\tilde g}=500$~GeV.
(iii) All other decay modes are at most of order $10^{-2}$.

\begin{figure}
\setlength{\unitlength}{1mm}
\begin{center}
\begin{picture}(160,60)
%
%
\put(5,-5){\mbox{
   \epsfig{figure=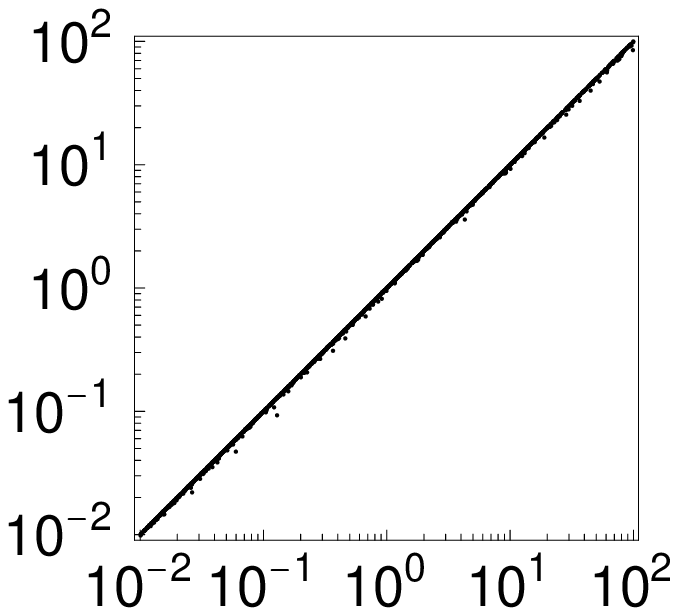,height=6.5cm,width=7.cm}}}
\put(3,5){\makebox(0,0)[br]{
    \begin{rotate}{90}
      {{\small $BR(\tilde g \rightarrow e \bar{b} t )
                        /BR(\tilde g \rightarrow \mu \bar{b} t)$}}
    \end{rotate}}}
\put(59,-6){\makebox(0,0)[br]{{\small $(\epsilon_1 / \epsilon_2)^2$}}}
\put(22,49){\mbox{{\small (a) }}}
%
%
\put(88,-5){\mbox{
   \epsfig{figure=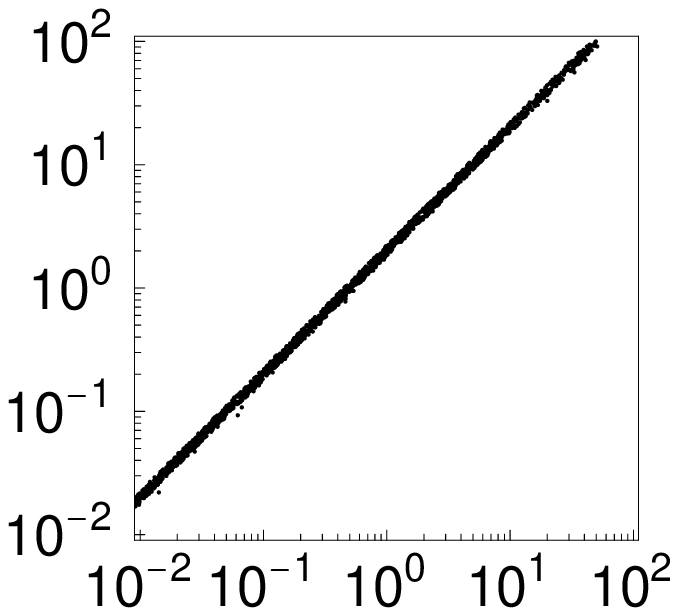,height=6.5cm,width=7.cm}}}
\put(86,5){\makebox(0,0)[br]{
    \begin{rotate}{90}
      {{\small $BR(\tilde g \rightarrow e \bar{b} t )
                        /BR(\tilde g \rightarrow \mu \bar{b} t)$}}
    \end{rotate}}}
\put(152,-6){\makebox(0,0)[br]{{\small $\tan^2\theta_{sol}$}}}
\put(105,49){\mbox{{\small (b) }}}
\end{picture}
\end{center}
\caption[]{a) Ratio 
$BR(\tilde g \rightarrow e \bar{b} t) /BR(\tilde g \rightarrow \mu \bar{b} t)$
as a function of $(\epsilon_1 / \epsilon_2)^2$; b) Ratio 
$BR(\tilde g \rightarrow e \bar{b} t) /BR(\tilde g \rightarrow \mu \bar{b} t)$
as a function of $(\tan \theta_{sol})^2$.}
\label{fig:gluinosb}
\end{figure}

The important decay modes for testing correlations between gluino
branching ratios and neutrino angles are the final states $l^\pm b t$
($l=e,\mu,\tau$). The sum of these decay modes is normally of the
order of a few per--cent.  In these decays the same class of couplings
induced by the effective vertex $\hat
L_i \hat T_L \hat B^c_R$ being proportional to $h_b \epsilon_i$
are probed as in the decays of the lighter stop
\cite{Restrepo:2001me} and in the decays of the lighter sbottom. As
can be seen in Figs.~\ref{fig:gluinosb}a and b, there is a clear
relation between the ratios of the final states into $l_i t b$ and the
corresponding ratio of $\epsilon_i$. In case of the ratio BR($\tilde g
\to e t b$) / BR($\tilde g \to \mu t b$) this implies a clear
correlation with the solar mixing angle as can be seen in
\fig{fig:gluinosb}b.

\subsection{Gravitino}

In gauge mediated SUSY breaking \cite{gmsb2} the gravitino $\tilde G$ is 
the LSP. Depending on the scale of SUSY breaking, its mass is typically 
of the order eV up to several MeV. If R-parity is broken, the gravitino 
decays, but it is long lived from the point of view of collider physics, 
because the width is proportional to an R-parity violating coupling squared 
and the ratio $(m_{\tilde G}/m_{SUSY})^4$. This implies that the gravitino
will escape the detector before decaying. 

However, one can consider the decays of the next-to-lightest SUSY particle 
(NLSP). For a gravitino with a mass of 1 eV one finds a partial width
for NLSP decays into gravitino of  O($10^{-3}$) eV and significantly 
smaller values for larger gravitino masses. Partial widths of the lightest 
neutralino (or chargino) into R-parity violating final state are of the same 
order of magnitude. This implies that branching ratios for R-parity 
violating final states of the neutralino are at least of $O(10^{-2})$. 
For slepton NLSPs typical partial widths into R-parity violating
states are of $O(1)$ eV in case of staus and thus clearly exceed 
the decay into a gravitino. Very similar arguments apply to all the 
other NLSP candidates discussed in the previous sections. 
Therefore, we conclude that in GMSB
models with bilinear R-parity violating terms, the decays of the NLSP
can be used to establish the correlations with neutrino physics 
we have discussed. 

\section{Summary}

We have calculated the decay patterns of various possible LSPs 
within bilinear R-parity violating supersymmetry. The main 
conclusion of the present work is that whichever SUSY particle 
is the LSP, measurements of branching ratios at future accelerators 
will provide a definite test of bilinear R-parity breaking as 
the model of neutrino mass. In case of GMSB, where the gravitino
is the LSP, we find that correlations with neutrino physics exist 
for the decays of the NLSP. 

One can state the above more carefully. Observation of a decaying 
LSP would provide proof that R-parity is violated. Measuring ratios 
of branching ratios then presents the ultimate cross-check of the 
completeness and uniqueness of the bilinear model. The most robust 
predictions of BRpV are shown in this paper, but many ratios of 
branching ratios are tightly constrained by neutrino physics. In 
fact, for several different LSP candidates many different decay 
channels have sizable branching ratios which should follow the 
specific patterns discussed in the previous sections. Thus this 
simplest model of R-parity violation can be over constrained by the 
measurements we have discussed and - in this sense - easily ruled out.

\section*{Acknowledgments}
  
This work was supported in part by the Spanish grant BFM2002-00345 
and by the European Commission RTN network HPRN-CT-2000-00148.  
M. H.  is supported by a Spanish MCyT Ramon y Cajal contract.
W.~P.~is supported by the 'Erwin Schr\"odinger fellowship No.~J2272' 
of the `Fonds zurF\"orderung der wissenschaftlichen Forschung' of 
Austria and partly by the Swiss `Nationalfonds'.

\end{document}